# Block Error Rate and UE Throughput Performance Evaluation using LLS and SLS in 3GPP LTE Downlink

Ishtiaq Ahmad, Zeeshan Kaleem, and KyungHi Chang

Electronic Engineering Department, Inha University

Ishtiaq001@gmail.com, zeeshankaleem@gmail.com, and khchang@inha.ac.kr

## Abstract

This paper provides a link and system level study of the downlink using OFDMA techniques based on LTE systems. At the link level, the Block Error Rate (BLER) performance curves are generated and 10 % points of BLER curves, calculated under SISO AWGN channel, are considered as reference to generate the SNR-to-CQI mapping. These mapping results are used in system level simulator (SLS) via MIESM link to system interface. Then the UE throughput is evaluated by considering proportional fair (PF), round robin (RR), and best CQI schedulers at different values of UE throughput CDF. In this paper we analyze that the PF scheduler outperforms other schedulers in fairness and average throughput, while the best CQI scheduler performs good under high SINR region. The RR is the best choice when resource allocation is to be done without considering any kind of channel condition.

## I. Introduction

The Long Term Evolution (LTE) is designed to meet the optimal data rate (data throughput) and efficient use of spectrum under the specified standards of 3rd Generation Partnership Project (3GPP) in Release 8. LTE is the evolutionary move based on UMTS standards including many changes and significant improvements identified by the 3GPP consortium. The goal of LTE to achieve the targets for downlink (DL) and uplink (UL) peak data rate requirements were set to 100 Mbit/s and 50 Mbit/s, respectively under the 20 MHz spectrum. These targets must be supported at low speed Up to 15 km/h and high speed up to 350 km/h for 5 km coverage and 100 km ranges, respectively. Moreover, in order to cope with spectrum fragmentation, the spectrum allocation of the system is flexible. The main reason for these changes in the Radio Access Network (RAN) system design is the need to provide higher spectral efficiency, lower delay, and more multi-user flexibility than the currently deployed networks. The LTE DL air interface is based on orthogonal frequency division multiple access (OFDMA), which converts the wide band frequency selective channel into set of many flat fading sub-channels to improve the spectral efficiency. In order to achieve the challenging spectral efficiency and user throughput targets, multiple antenna schemes, turbo coding and fast link adaptation are also included in the specifications. In order to optimize the system data throughput and the coverage area for a given transmission power, LTE make use of the Adaptive Modulation and Coding (AMC). By using AMC, in the existence of interference level from other cells and noise level at the receiver, the transmitter assign the data rate for each user depending on the channel quality from the serving cell. In addition, OFDMA allows for frequency domain scheduling; generally assign good sub-channels to the individual users. The scheduler assign proper allocation of the resources with certain objectives like required QoS, high cell throughput under existing channel condition, Fairness among UEs & application, Load balancing among the users and limiting the impact of interference of cell edge users.

To study the performance of LTE systems the MATLAB based open source downlink link level simulator (LLS) [1] [2] and SLS [3] have been developed. The goal of open source simulators is to facilitate comparison with work of different research groups. The link level simulations emphasize mainly on BLER vs SINR ,adaptive coding and modulation or physical layer modeling for system level [4], while system level simulations focus more on network-related issues such as scheduling [5], mobility handling or interference management [6]. In this paper, the BLER of SISO in 20 MHz LTE System for different Modulation and Coding Scheme (MCS) are investigated in terms of SNR using link level LTE Simulator. The UE throughput performance based on different scheduling schemes is investigated by using LTE System Level Simulator.

The rest of the paper is organized as: In Section II, we present the system and channel model used in the simulation, while in the section III the LL & SL performance is being discussed. The simulations and discussions on results are presented in Section IV. Finally, we conclude our paper in Section V.

## II. System and Channel Model

The OFDM signal has a time and frequency domains. In the time domain, the LTE signal is composed of successive frames. Each frame has duration of 10 ms and it is divided into ten equally 1 ms long sub-frames. Each sub-frame consists of two equally long slots with 0.5 ms time duration. For normal cyclic prefix length each slot consists of Ns = 7 OFDM symbols. In the frequency domain, the OFDM technique converts the LTE wide band signal into a number of narrowband signals. Each narrowband signal is transmitted on one subcarrier frequency. In LTE the spacing between subcarriers is fixed to 15 KHz. Twelve adjacent subcarriers, occupying a total of 180 KHz, of one slot forms the so-called Resource Block (RB). The number of Resource Blocks in an LTE slot depends on the allowed system bandwidth. We used the maximum allowed BW (20 MHz) in LTE, where the maximum 100 RBs are allowed.

Table1. LL and SL simulation parameters

| Parameters | Values |
|---|---|
| Link Level Parameters | |
| Bandwidth | 20 MHz |
| Number of Sub-carriers per PRB | 12 |
| Maximum Number of PRBs | 100 |
| Channel Coding | Turbo Code (1/3) |
| AMC formats | QPSK: 1/3, 1/2, 2/3, 4/5<br>16QAM: 1/2, 2/3, 4/5<br>64QAM: 2/3, 4/5 |
| Antenna Scheme | SISO (1x1) |
| Channel Model | AWGN |
| System Level Parameters | |
| Bandwidth | 20 MHz |
| Antenna Scheme | 2 x 2 CLSM |
| Channel Model | WINNER II |
| Traffic Model | Full Buffer |
| Effective SNIR Mapping | MIESM |
| UE Speeds | 5 km/hr |
| Number of UEs | 20 UEs per cell |

The LLS simulations are performed under the AWGN channel model. The details of the main parameters used for the LLS and SLS are given in the Table 1. In case of SLS we used the WINNER II as the channel model and 5km/h mobility for the users is considered.

## III. Simulator Structure in 3GPP LTE

### 3.1 Structure of Link Level Simulator

The LTE LLS consists of the following functional blocks: transmitter (eNB), AWGN Channel Model & receiver (UE).The details are given in the block diagram given in the figure 1.

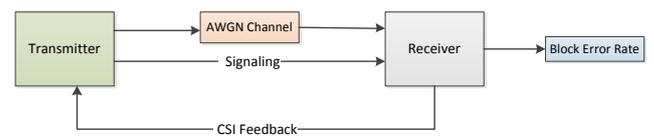

Figure 1. Link level structure.

In SISO OFDM systems, the maximal UE Capacity depends upon the available bandwidth and the parameter of the OFDM signal, like the number of subcarriers and the modulation order (QPSK, 16QAM, 64QAM). For a given bandwidth (BW) the maximal data capacity on LLS side in bits per second can be approximated by the following equation;

$$\text{Max UE Capacity} = BW * CR_{eff} * (1 - BLER) \quad (1)$$

Where BW is the total bandwidth offered by LTE, $CR_{eff}$ is the effective code rate of the selected modulation scheme and BLER is the block error rate.

### 3.1.1 Link-to-System Level Interface (L2SI)

The L2SI are designed to allow a high accuracy on the exchange of metrics between LL & SL Simulators. The numbers of methods are used in the literature like Effective

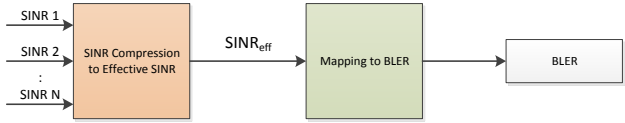

Figure 2. Link-to-system level interface.

Exponential SINR Mapping (EESM) & Mutual Information Effective SINR Mapping (MIESM) to calculate the average SINR values. In our simulations we used the MIESM because its performance is better than other schemes. The procedure to calculate the MIESM is shown in the figure 2.

The equation 2 is used to calculate the MIESM and is given as;

$$SNR_{eff} = \alpha_1 * I^{-1}\left( \frac{1}{P} \sum_{P=1}^{P} \left( I\frac{SINR_P}{\alpha_2} \right) \right) \quad (2)$$

After calculating these mapping parameters, the look up tables (LUT) is used in the SLS to save the time by reducing the computational complexity.

### 3.2 Structure of System Level Simulator in 3GPP LTE

The performance of whole network is analyzed by SLS; the network consists of multiple eNBs that covers the specific area in which multiple UEs are supported or moving around. The LLS are suitable for developing receiver structures, coding schemes or feedback strategies, it is not possible to reflect the effects of issues such as cell planning, scheduling, or interference using this type of simulations. Simulating the totality of the radio links between the UEs and eNBs is an impractical way of performing SLS due to the vast amount of computational power that would be required. Thus, in SLS the physical layer is abstracted by simplified models that capture its essential characteristics with high accuracy and simultaneously low complexity. Similarly to other SLS, the core part consists of: (i) a link measurement model [7] and (ii) a link performance model [8]. The link measurement model reflects the link quality, given by the UE measurement reports, and is required to carry out link adaptation and resource allocation while the link performance model predicts the BLER of the link, based on the receiver SINR and the transmission parameters (e.g., modulation and coding) [9].

The link performance model determines the BLER at the receiver given a certain resource allocation and Modulation and Coding Scheme (MCS). For LTE, 15 different MCSs are defined, driven by 15 Channel Quality Indicator (CQI) values. To assess the BLER of the received Transport Blocks (TBs), a set of Additive White Gaussian Noise (AWGN) link-level performance curves are discussed in the results section. The SINR-to-BLER mapping then requires of an effective SINR value that can be obtained as discussed in section 3.1.1.

## IV. Simulation Results

### 4.1 Block Error Rate Performance by LLS

To obtain the Block Error Ratio (BLER) for MCS corresponding to each CQI value, AWGN simulations were performed. The MCS determines both the modulation and coding schemes.

Figure 3 shows the BLER results of CQIs 1-15 without using HARQ. Each curve is spaced approximately 2 dB from each other.

The CQI feedback report is calculated based on the Sub-carrier SINRs and the target transport BLER. The CQI reports are generated by an SINR-to-CQI mapping and made available to the eNB implementation via a feedback channel with adjustable delay. At the transmitter, the appropriate MCS is selected by the CQI to achieve the targeted BLER during the transmission. In the link performance model, an AWGN-equivalent SINR is obtained via MIESM.

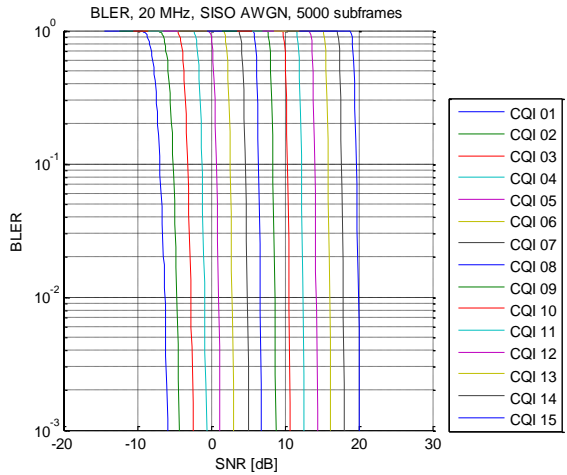

Figure 3. BLER curves under SISO AWGN channel for 15 CQI values.

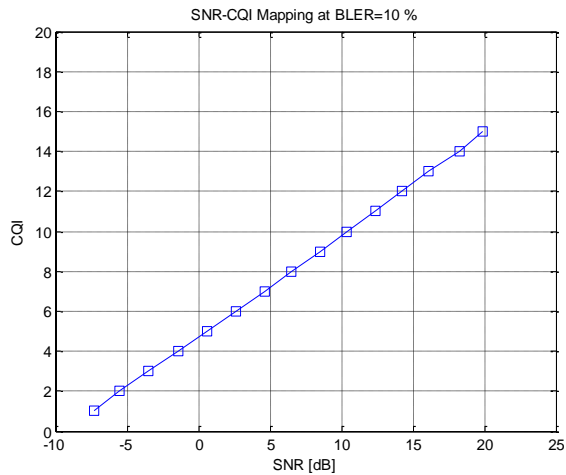

Figure 4. SNR-CQI mapping at BLER =10 % for 20 MHz BW.

The defined CQIs use coding rates between 1/13 and 1 combined with 4-QAM, 16-QAM and 64-QAM modulations. For the CQI feedback strategy, the SINR-to-CQI mapping is realized by taking the 10% points of the BLER curves, to obtain the mapping as shown in Figure 4.

### 4.2 UE Throughput Evaluation in SLS

The SLS [3] is used by setting some of the simulation parameters as given in Table 1. The simulation environment consists of just only one tier. The users are deployed randomly in the center cell. The other cells have only eNBs which provides the interference to the UEs in the center cell.

The throughput performance of users is compared by allocating the resources by using different scheduling schemes such as round robin (RR), best CQI and proportional fair (PF) scheduling. The figure 5 depicts the comparative throughput performance of UEs under these schedulers. We compare the throughput for UEs at three different points of CDF. At 50 % of CDF, the best CQI have the worst performance because almost all UEs are in the outage stage, while the RR and PF scheduler served fairly to the UEs with an acceptable throughput of 2 Mbps and 2.9 Mbps respectively. The PF scheduler is best than the RR scheduler in serving around 90 % because it considers the fairness among the UEs. At 95 % of CDF the PF performance degrades as compared to RR because PF also considered the hungriness of the UEs i.e depends upon the resources previously being allocated to the same UE.

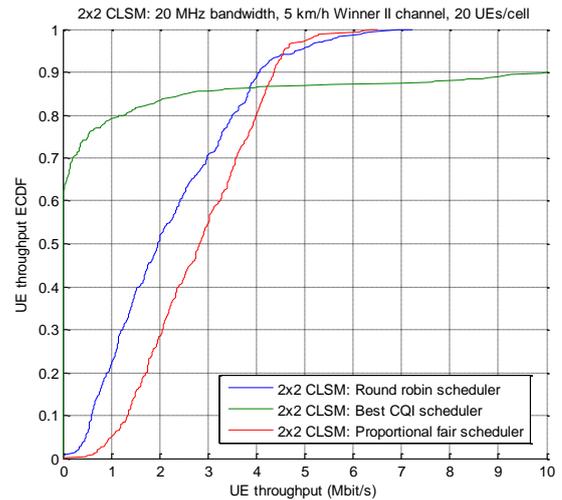

Figure 5. UE throughput CDF: 2x2 CLSM, round robin, max C/I, and PF scheduler.

The 100 % UE get the maximum throughput of 7 Mbps by reaching the limiting point in the case of PF and RR, while at the same throughput the best CQI scheduler is still assigning the resources to the UEs and even did not reach its limiting point at 10 Mbps. So if the UEs in focus are with the best SINR conditions, then the best choice to be considered is best CQI scheduling algorithm.

The round robin is very simple to implement and giving the best fairness. But it would offer poor performance in terms of UE throughput as compared to the PF scheduler. The PF offers high UE throughput as well as fairness satisfactorily. Thus, Proportional Fair (PF) scheduling may be the best option when focusing on both the fairness and UE throughput.

## V. Conclusions

This paper elaborates the LL and SL performance of LTE systems. The LLS results obtained from BLER curves at 10 % are used for SNR-CQI mapping in SLS. Then the performance of PF, RR, and best CQI scheduling algorithms are compared in SLS. The results show that at 50 % of CDF, PF outperforms all the schedulers in the aspect of UEs throughput. The CQI scheduler is the only choice when the channel condition is the best without considering any kind of fairness among the UEs. The RR is considered to distribute the resources equally among users and without caring the effect of channel conditions.

## ACKNOWLEDGMENT

This work was supported by the IT R&D program of MKE/KEIT [10039568, Development of Policy-based Load balancing LTE-WiFi Dual Mode Femtocell].